\documentclass[conference,english,final,svgnames]{IEEEtran}

\usepackage{hyperref}
\usepackage{commands}
\usepackage{microtype,babel,csquotes}
\usepackage{stmaryrd}
\usepackage[style=ieee, doi=false, isbn=false, url=false]{biblatex}
\usepackage{xcolor}

\IEEEoverridecommandlockouts

\hypersetup{
    pdftitle={Convergence Guarantees for Unmixing PSFs over a Manifold with Non-Convex Optimization},
    pdfauthor={Santos Michelena, Maxime Ferreira Da Costa, José Picheral},
}

\title{Convergence Guarantees for Unmixing PSFs over a Manifold with Non-Convex Optimization}

\author{
    \IEEEauthorblockN{
        Santos Michelena\IEEEauthorrefmark{1}\IEEEauthorrefmark{2}, Maxime Ferreira Da Costa\IEEEauthorrefmark{1}, and José Picheral\IEEEauthorrefmark{1}
    }
    \IEEEauthorblockA{
        \IEEEauthorrefmark{1}Laboratory of Signals and Systems, CentraleSupélec, Université Paris–Saclay, CNRS, Gif-sur-Yvette, France\\
        \IEEEauthorrefmark{2}Iumtek, Orsay, France
    }
    \vspace{-20pt}
  \thanks{
       MF acknowledges support from ANR funding ANR-24-CE48-3094. Emails:  \{santos.michelena, maxime.ferreira, jose.picheral\}@centralesupelec.fr.
    }
}
    
\begin{document}

\maketitle

\begin{abstract}
    The problem of recovering the parameters of a mixture of spike signals convolved with different PSFs is considered. Herein, the spike support is assumed to be known, while the PSFs lie on a manifold. A non-linear least squares estimator of the mixture parameters is formulated. In the absence of noise, a lower bound on the radius of the strong basin of attraction \emph{i.e.}, the region of convergence, is derived. Key to the analysis is the introduction of coherence and interference functions, which capture the conditioning of the PSF manifold in terms of the minimal separation of the support. Numerical experiments validate the theoretical findings. Finally, the practicality and efficacy of the non-linear least squares approach are showcased on spectral data from laser-induced breakdown spectroscopy.
\end{abstract}

\section{Introduction}

Point spread function (PSF) unmixing is the problem of reconstructing multiple groups of spikes after their convolution with a group-dependent PSF.

Mixtures of sparse signals typically arise whenever the experimenter records a superposition of phenomena produced by different modalities. This parametric signal model is encountered in various areas of applied science and engineering, such as in super-resolution imaging~\cite{huang2008three,huang2018single}, impulse sorting from neural recordings~\cite{knudson2014inferring,li2014sparse}, multi-path channel identification in wireless communication~\cite{applebaum2012asynchronous,chi2013compressive}, or calibration-free spectroscopy~\cite{HU2022116618}, among others.

While the blind deconvolution problem~\cite{li2016identifiability,li2019multichannel}---which assumes a \emph{unique} modality---is well-studied in signal processing,  the state-of-the-art in resolving sparse mixtures of unknown PSFs is limited. In~\cite{romberg2010sparse}, linear programming is first investigated when the spikes are assumed on a grid. Other approaches relying on the atomic norm minimization framework~\cite{chi2020harnessing} remove the grid assumption and achieve stable reconstruction of both the signals and the PSFs~\cite{chi_guaranteed_2016, Li_2020}. Yet, the PSFs are constrained to lie in a predetermined known low-dimensional subspace, which may be impractical. Furthermore, the computational complexity of convex-based approaches can be prohibitive in real-world applications.

On the other hand, non-convex methods have successfully recovered sparse signals under strong structural assumptions in a scalable manner~\cite{traonmilin2020basins,traonmilin2024strong}. When measurements are taken in the Fourier domain, non-convex optimization is shown to deconvolve well-separated sparse signals~\cite{costa_local_2023, gabet2025global}.

In \cite{LI2019893}, the authors investigate the local geometry of blind deconvolution using non-linear least squares minimization to estimate both the PSF and the signal of interest from a single measurement and provide an estimate for the radius of the strong basin attraction of their method. However, the approach assumes the single modality is in a low-dimensional subspace. Overall, the current techniques are maladapted to many applications, such as laser-induced breakdown spectroscopy (LIBS), where the shapes of the PSFs depend on the complex non-linear physical interactions within the laser-generated plasma~\cite{mars-spectrum-fitting}.

\subsection{Contributions and Organization of the Paper}

Motivated by application to LIBS, where accurate estimation of the plasma parameters is the crux to recover the concentration of chemical species~\cite{HU2022116618}, we consider the problem of unmixing spikes with \emph{known} support---corresponding to emission wavelengths---and relax the stringent low-dimensional subspace assumption by assuming an ensemble of PSFs lying on a postulated \emph{manifold}. We formulate a non-linear least squares estimator to recover the amplitudes of the spikes and the shape parameters of the PSFs composing the recorded signal. In the absence of noise, a lower bound on the size of the region of strong convexity around the ground truth is established in terms of \emph{minimum separation} of the support, as well as novel  \emph{coherence and interference functions} characterizing the conditioning of PSF manifold. Numerical experiments on synthetic and LIBS data showcase our approach's effectiveness.

The rest of the article is organized as follows. Our mixture recovery problem is introduced in Subsection~\ref{subsec:problem_formulation}. Section~\ref{sec:main_results} recalls the importance of strong basins of attraction in non-convex optimization and presents our key novel coherence and interference metrics for the PSF manifold. A lower bound on the radius of the basin of attraction of the ground truth parameters is established in terms of those metrics in Theorem~\ref{thm:main-result}. A sketch proof of Theorem~\ref{thm:main-result} is drawn in Section~\ref{sec:proof}. Section~\ref{sec:simulations} conducts numerical simulations validating our theory, and presents an application to aluminum concentration estimation on real-world LIBS data. Finally, a conclusion is drawn in Section~\ref{sec:conclusion}.

\subsection{Problem Formulation}\label{subsec:problem_formulation}

We assume a parametric class of PSFs $g(\theta,\cdot)$, where $\theta \in \Theta \subset \mathbb{R}$ is a univariate shape parameter. The investigated problem is to recover a signal of the form
\begin{align}
\label{eq:signal-model}
    t \mapsto x(t) = \sum_{i = 1}^p \sum_{\ell = 1}^{L_i}{\eta^\star_{i,\ell}} g(\theta_i^\star, t - t_{i,\ell})
\end{align}
from $N$ samples taken uniformly over an the interval $I = [-\tfrac{T}{2}, \tfrac{T}{2}]$, for some $T>0$, and at the timestamps $\{u_s = -\tfrac{T}{2} + \tfrac{T(s-1)}{N-1}\}_{s=1}^N$. In our setting, both the source locations $\{t_{1,1}, \dots, t_{1,L_1}, \dots, t_{p,1}, \dots, t_{p,L_p}\}$ and the parametric class $ g(\cdot, \cdot)$ are assumed to be \emph{known}; While the true shape parameters $ \{\theta_1^\star, \dots, \theta_p^\star\} $ and coefficients $\{\eta_{i,\ell}^\star\}$ remain \emph{unknown} at the time of observation. Furthermore, the mixture number $p$ and the
number of spikes per modality $\{L_1, \dots, L_p\}$ are \emph{known}. 
Let $\bg_{\ell,i} = [g(\theta_i, u_s - t_{i,\ell})]_{s=1}^N$, and define the map $\btheta \mapsto\bG(\btheta)$
\begin{align}
	\bG(\btheta) = \begin{bmatrix}
        \bg_{1,1} & \dots & \bg_{1,L_1} & \dots & \bg_{p,1} & \dots & \bg_{p,L_p}
     \end{bmatrix}.
\end{align}
Furthermore, we define
\begin{equation}\etab^{\star} = \begin{bmatrix}
    \eta^{\star}_{1,1} & \dots & \eta_{1, L_1}^{\star} & \dots & \eta_{p,1}^{\star} & \dots & \eta_{p, L_p}^{\star}
\end{bmatrix}^\top \in \R^{\overline{L}},
\end{equation}
where $\overline{L} = \sum_{i=1}^p L_i$. The noiseless measurements $\bx^\star$ read
\begin{equation}\label{eq:ground_truth}
    \bx^{\star} = \bG(\btheta^{\star})\etab^{\star}.
\end{equation}

We study the problem of estimating the parameters $\{\bm{\theta}^\star, \bm{\eta}^\star\}$ in~\eqref{eq:ground_truth} by solving the non-linear least-squares program
\begin{align}
    \label{eq:problem}
    \argmin_{\btheta, \etab} \left\{ \L(\btheta, \etab) = \frac{1}{2}\norm{\bG(\btheta)\etab- \bx^{\star}}_2^2 \right\}.
\end{align}
 For our analysis, we assume $g(\cdot, \cdot)$ is twice continuously differentiable with respect to its first argument $\theta$, meaning that $g(\cdot, t) \in \mathcal{C}^2(\Theta)$ for all $t \in I$, and write $\partial_a g(\cdot, \cdot)$ the $a$-th derivative with respect to $\theta$. Furthermore, the PSFs and their derivatives up to the second order are assumed both integrable and square-integrable over $\mathbb{R}$, that is
\begin{align}
     \forall \theta \in \Theta,\; \forall a \in \{0,1,2\} ,\quad \partial_a g(\theta, \cdot) \in L^1(\mathbb{R}) \cap L^2(\mathbb{R}).
\end{align}
 Additionally, we require that $g(\theta, \cdot)$ be even for all $\theta \in \Theta$.

\section{Main Results}\label{sec:main_results}

\subsection{Strong Basins of Attraction}

It is clear that the ground truth parameters $(\btheta^\star, \etab^\star)$ are a solution of~\eqref{eq:problem} as $\mathcal{L}(\btheta^\star, \etab^\star) = 0$. However, given the non-convex nature of our loss function~\eqref{eq:problem}, bad local minima may exist. Herein, we investigate the \emph{local convergence} of line search methods such as gradient descent and the method of Gauss-Newton in a neighborhood of $(\bm{\theta}^\star, \bm{\eta}^\star)$. To that end, we seek a characterization of the strong basin of attraction of the loss~\eqref{eq:problem} around the ground truth parameters.

That is, a neighborhood $\mathcal{N}$ of $(\btheta^\star, \etab^\star)$ such that there exist two constants $0 < \xi \leq \gamma < \infty$ such that the loss~\eqref{eq:problem}  be $\xi$-strongly convex and $\gamma$-smooth, \emph{i.e.} if $(\btheta, \etab) \in \mathcal{N}$ then
\begin{align}
    \label{eq:min-eigval}
     \forall \bu, \; \xi \norm{\bu}_2^2 \leq \bu^{\top} \nabla^2 \mathcal{L}(\btheta, \etab)\bu \leq  \gamma \norm{\bu}_2^2.
\end{align}

Line search methods are guaranteed to converge to the global solution $(\btheta^\star, \etab^\star)$ when initialized within the strong basin of attraction $\mathcal{N}$. However, the rate of convergence varies depending on the method. For example, it is well established in the literature that the method of steepest descent will converge linearly whenever the step size $\alpha$ is chosen as $\alpha = \gamma/\xi$ (see \emph{e.g.} \cite[Theorem 2.1]{Nocedal2006}).

\subsection{Coherence and Interference Functions}
\label{sec:coherence}

In this section, we introduce key metrics to our analysis of the radius of the basin~\eqref{eq:min-eigval}. The \emph{minimal separation} of the support $\Delta$ is known to affect the well-posedness of sparse inverse problems~\cite{ferreira2020stable, hockmann2023weak} and is defined in our setting as
\begin{equation}
    \Delta \coloneqq \min_{i,j} \min_{(i,\ell) \neq (j,k)} |t_{i,\ell} - t_{j,k}|.
\end{equation}
For $a \in \{0,1,2\}$, we let the vectors
\begin{equation}
    \partial_a \bg(\theta, t) := \left[ \partial_a g(\theta, u_s - t) \right]_{s=1}^N \in \R^N.
\end{equation}
We introduce the \emph{coherence}, which characterizes the maximum correlation between the dictionary elements and their derivatives when separated by at least $\Delta$, and is defined by
\begin{equation}\label{eq:def_coherence}
    \mu_{a,b}(\theta_i, \theta_j, \Delta) 
    \coloneqq \sup_{|\delta|\geq \Delta} \left| \partial_a \bg(\theta_i, 0)^{\top} \partial_b \bg(\theta_j, \delta)  \right|.
\end{equation}

The \emph{coherence function} between $\partial_a g(\theta_i, \cdot)$ and $\partial_b g(\theta_j, \cdot)$ writes
\begin{align}
\label{eq:def_coherence_function}
\MoveEqLeft \mathcal{C}_{a,b}(\theta_i, \theta_j, \Delta) & \nonumber \\
    & \hspace{-16pt}:= \begin{cases}
        2\sum\limits_{m \in \N \setminus \{0\}} \mu_{a,b}(\theta_i, \theta_j, m\Delta) & \text{if } i=j\text{ and }a = b \\
        2\sum\limits_{m \in \N } \mu_{a,b}(\theta_i, \theta_j, m\Delta) & \text{otherwise.}
    \end{cases}
\end{align}
Similarly, we define \emph{interference functions} to capture the decay rate of $g(\theta, \cdot)$ and its derivatives when the support is separated by at least $\Delta$, which we define as
\begin{equation}\label{eq:def_interference_functions}
    \I_a(\theta_i, \Delta) \coloneqq \left|\partial_ag(\theta_i, 0)\right| +  2 \sum_{m \in {\N \setminus \{0\}}} \sup_{|\delta| \geq m\Delta}|g_a(\theta_i, \delta)|.
\end{equation}

\begin{figure}[t]
    \centering
    \includegraphics[width=\linewidth]{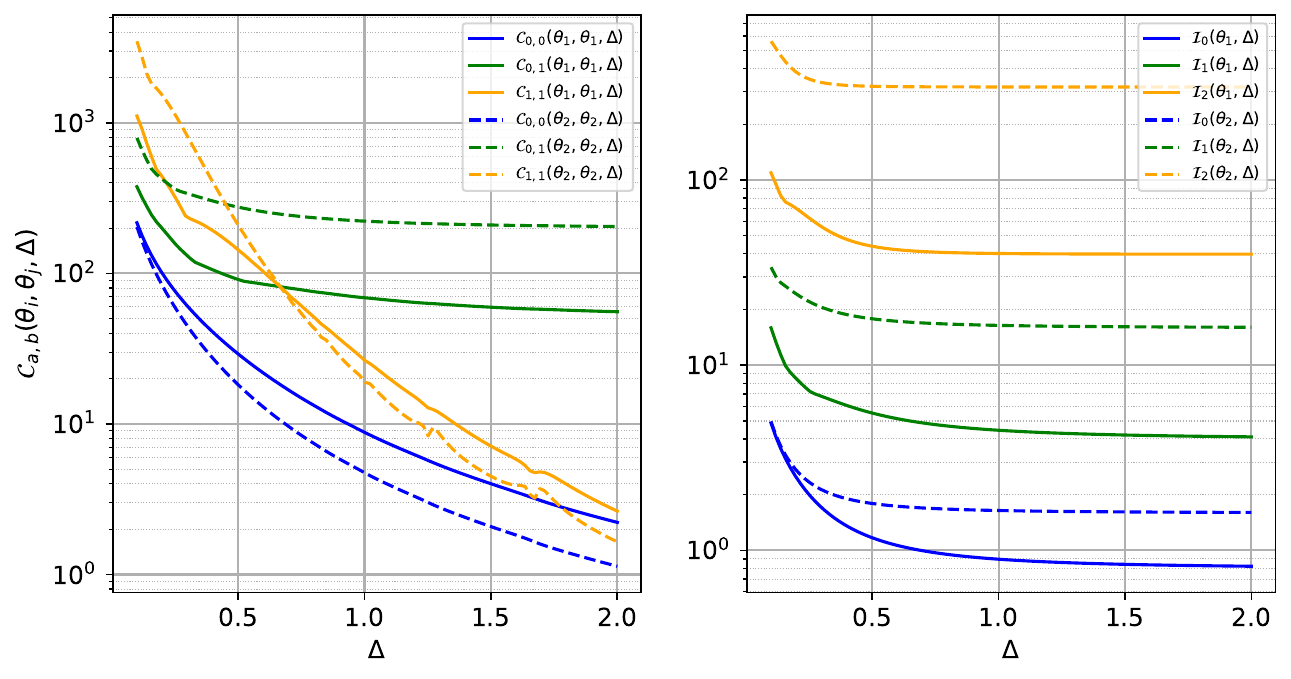}
    \vspace{-12pt}
    \caption{Graphs of the coherence functions (left) and the interference functions (right) for the class of Lorentz PSF $g(\theta, t) = \theta / (\pi (\theta^2 + t^2))$ in the range $\Delta \in [0.1, 2]$ for the parameters $\theta_1 = 0.2$ and $\theta_2=0.1$.}
    \label{fig:coherence}
\end{figure}

For illustration purposes, Figure~\ref{fig:coherence} pictures the coherence and interference functions for the class of Lorentz kernels. Of particular importance in the sequel, we remark that the coherence functions $C_{0,0}(\theta_i, \theta_i, \Delta)$ and $C_{1,1}(\theta_i, \theta_i, \Delta)$ tend to $0$, while $C_{1,0}(\theta_i, \theta_j, \Delta)$ converges to a constant value as $\Delta$ increases. This property is consistent with Equation~\eqref{eq:def_coherence} and $\mu_{a,b}(\theta_i, \theta_j,\Delta) \to 0$ as $\Delta \to \infty$.  Similarly, the interference functions satisfies $\mathcal{I}_a(\theta_i,\Delta) \to \left|\partial_a g(\theta_i, 0)\right|$ with $\Delta \to \infty$.

\subsection{Main Result Statement}
Denote by $\mathcal{B}_\varepsilon = \{(\btheta, \etab): \ \norm{\btheta - \btheta^{\star}}_{\infty} + \norm{\etab - \etab^{\star}}_{\infty} < \eps\}$ 
and let the shorthands $\eta_{i,\min}^\star = \min_\ell \{\eta_{i,\ell}^\star\}$ and $\eta_{i,\max}^\star = \max_\ell \{\eta_{i,\ell}^\star\}$. As our main result, we establish that $\mathcal{B}_\varepsilon$ is contained in the strong basin of attraction of the ground truth for the loss function~\eqref{eq:problem} whenever $\varepsilon$ is small enough. Furthermore, we express the convexity and smoothness parameters~\eqref{eq:min-eigval} as a function of the coherence and interference functions introduced in Section~\ref{sec:coherence}.

\begin{theorem}
\label{thm:main-result}
    Assume the maps $\theta \mapsto \mathcal{C}_{a,b}(\theta, 
    \theta',\Delta)$ and $\theta \mapsto \mathcal{I}_{a}(\btheta,\Delta)$ are Lipschitz with constant $C_\Delta$, and the map $\btheta \mapsto \bG(\btheta)$ is Lipschitz in the $\ell_\infty$-topology with constant $K$. Let
    \begin{subequations}
        \begin{align}
        c^\star_- &\coloneqq \frac{1}{2} \min_{i\in \llbracket p \rrbracket} \left\{\eta^{\star^2}_{i,\min} \mu_{1,1}(\theta_i^
        \star, \theta_i^\star, 0), \mu_{0,0} (\theta_i^\star, \theta_i^\star, 0) \right\} \label{eq:def_c_minus}\\
        c^\star_+ &\coloneqq \frac{3}{2} \max_{i\in \llbracket p \rrbracket} \left\{\eta^{\star^2}_{i,\max} \mu_{1,1}(\theta_i^
        \star, \theta_i^\star, 0), \mu_{0,0} (\theta_i^\star, \theta_i^\star, 0) \right\} \label{eq:def_c_plus}\\
        r^\star &\coloneqq \max_{i\in \llbracket p \rrbracket} \Big\{ \sum_{j=1}^p \eta^\star_{j,\max}\C_{0,1}(\theta_i^\star,\theta_j^\star, \Delta) + \C_{0,0}(\theta_i^\star, \theta_j^\star, \Delta) , \nonumber \\
        &  \eta^{\star}_{i,\max}  \sum_{j=1}^p \eta^{\star}_{j,\max}\C_{1,1}(\theta_i^\star,\theta_j^\star, \Delta) + \C_{1,0}(\theta_i^\star, \theta_j^\star, \Delta)   \Big\} \label{eq:def_r}\\
        q^\star &\coloneqq \max_{i\in \llbracket p \rrbracket} \left\{ \norm{\etab_i^\star}_{1}\I_2(\theta_i^\star, \Delta)) \hspace{-2pt}+\hspace{-2pt} L_i \I_1(\theta_i^\star, \Delta) \right\} \nonumber \\
        & \qquad \qquad + K\norm{\etab^{\star}}_{\infty} + 4pC_\Delta\max\{1, \norm{\etab^{\star}}_{\infty}^2 \}. \label{eq:def_q} 
    \end{align}
    \end{subequations}
    Furthermore, assume $r^\star < c^\star_-$, and select $\varepsilon \in [0, \tfrac{c^\star_- - r^\star}{q^\star})$. Then there exists $N_0 \in \mathbb{N}$ such that if $N \geq N_0$ then $\mathcal{B}_\varepsilon$ is a strong basin of attraction for the loss~\eqref{eq:problem} with constants
    \begin{align*}
        \xi &= c^\star_- - r^\star - q^\star\varepsilon; & \gamma&= c^\star_+ + r^\star + q^\star\varepsilon;
    \end{align*}
    in Equation~\eqref{eq:min-eigval}. Hence, any line search method converges given an initialization $(\btheta^0,\etab^0) \in \mathcal{B}_{\varepsilon_0}$ with $\varepsilon_0 = \frac{c^\star_- - r^\star}{q^\star}$.
\end{theorem}

The radius of the convexity region predicted by Theorem~\ref{thm:main-result} is driven by two antagonist factors.
On one hand, larger values of $c^\star_-$ and $c^\star_+$ in Equations~\eqref{eq:def_c_minus} and \eqref{eq:def_c_plus} widen the basin and are achieved by PSFs with larger energies $\mu_{0,0} (\theta_i^\star, \theta_i^\star, 0)$, or by PSFs whose first derivatives with respect to $\theta$ have larger energies $\mu_{1,1} (\theta_i^\star, \theta_i^\star, 0)$.
On the other hand, larger values of $r^\star$ from increased coherence functions or larger values of $q^\star$ from increased interference functions shrink the basin. From the definitions~\eqref{eq:def_coherence_function} and \eqref{eq:def_interference_functions}, lower coherence and interferences are obtained by increasing $\Delta$, or by considering a parametric class of PSF $g(\theta, t)$ with faster decays in the asymptotic $t\to \infty$.
 
\section{Proof Elements of Theorem~\ref{thm:main-result}}\label{sec:proof}

The proofs of intermediate lemmas are omitted due to space limitation. First, the Hessian $\nabla^2 \mathcal{L}(\btheta, \etab)$ of the loss~\eqref{eq:problem} writes
\begin{align}\label{eq:Hessian_expression}
    \MoveEqLeft \nabla^2 \mathcal{L}(\btheta, \etab) = \hspace{-2pt}\underbrace{\begin{bmatrix}
    \etab^{\top} & \bm{0}_{1, \overline{L}}  \\
    \bm{0}_{\overline{L}, 1} & \Id_{\overline{L}}
    \end{bmatrix}
    \begin{bmatrix}
    \bG_1^{\top}\bG_1 & \bG_1^{\top}\bG_0 \\
        \bG_0^{\top}\bG_1 & \bG_0^{\top}\bG_0
    \end{bmatrix}
    \begin{bmatrix}
    \etab & \bm{0}_{1, \overline{L}}  \\
    \bm{0}_{\overline{L}, 1} & \Id_{\overline{L}}
    \end{bmatrix}}_{\coloneqq \E} \nonumber \\
    & \qquad \qquad \qquad + \underbrace{\begin{bmatrix}
        \br^{\top} \bG_2\etab & \br^{\top}\bG_1 \\
        (\bG_1)^{\top}\br & \bm{0}_{\overline{L}, \overline{L}}
    \end{bmatrix}}_{\coloneqq \bm{R}},
\end{align}
where, for $a \in \{0,1,2\}$, we write $\bG^i_a := [\partial_a\bg_{i,1}, \cdots
\partial_a \bg_{i,L_i}]$ $\in \mathbb{R}^{N \times L_i}$, and $\bG_a = [\bG^1_a,\cdots, \bG^p_a] \in \mathbb{R}^{N \times \overline{L}}$. 
where $\br = \bG(\btheta) \etab - \bx^{\star}$ is the model residual. In the view of~\eqref{eq:min-eigval}, we aim to derive values $\xi$ and $\gamma$ in Theorem~\ref{thm:main-result} from a lower (resp. upper) bound of the spectrum from the expression of the Hessian matrix~\eqref{eq:Hessian_expression}.
To proceed, we denote by $\D$ the diagonal part of $\E$. Since $\nabla^2 \mathcal{L}(\btheta, \etab)$ is symmetric, its eigenvalues are real. One may bound the spectrum of $\nabla^2 \mathcal{L}(\btheta, \etab)$ with Weyl's inequality
 (see \emph{e.g.}~\cite[Theorem 8.1.5]{golub2013MatrixComputations}) as
\begin{subequations}\label{eq:weyl}
\begin{align}
    \hspace{-6pt} \lmin(\nabla^2 \mathcal{L}(\btheta, \etab)) &\geq \lmin(\D) - \norm{\nabla^2 \mathcal{L}(\btheta, \etab) - \D}_{\infty} \label{eq:lower_Weyl} \\
    \hspace{-6pt} \lmax(\nabla^2 \mathcal{L}(\btheta, \etab)) &\leq \lmax(\D) + \norm{\nabla^2 \mathcal{L}(\btheta, \etab) - \D}_{\infty} \label{eq:upper_Weyl}
\end{align}
\end{subequations}
Then, values of $\xi$ and $\gamma$ can be obtained by lower (resp. upper) bounding the right-hand side of~\eqref{eq:lower_Weyl} (resp. \eqref{eq:upper_Weyl}). This is achieved by controlling the three quantities $\lmax(\D)$, $\lmin(\D)$ and $\norm{\nabla^2 \mathcal{L}(\btheta, \etab) - \D}_{\infty}$ independently as functions of $\Delta$ and the error radius~$\varepsilon$.

The eigenvalues of $\bm{D}$ are controlled with Lemma~\ref{lem:spectrum_D}.
\begin{lemma}\label{lem:spectrum_D}
Under the assumptions of Theorem~\ref{thm:main-result}, one has
\begin{equation}\label{eq:bound_D}
        c_-^\star \leq \lmin(\bm{D}) \leq \lmax(\bm{D})  \leq c_+^\star.
\end{equation}
\end{lemma}
Next, the quantity $\norm{\nabla^2 \mathcal{L}(\btheta, \etab) - \D}_{\infty}$ is controlled by bounding the $\ell_{\infty}$-norm of each of the sub-block of $\norm{\nabla^2 \mathcal{L}(\btheta, \etab) - \D}_{\infty}$. As the matrices $\bG^i_a$ or $\bG_a^{i^{\top}}\bG_b^j$ for $a,b \in \{0,1\}$, and for all $i,j \in \{1,\dots, p\}$ are ubiquitous in the block decomposition, it is essential to control their $\ell_\infty$-norm. This is achieved in Lemma~\ref{lem:bound_G} by harnessing that for any $m\in \mathbb{N}\backslash\{0\}$, there exist at most two support points $\{t_{i,\ell}\}_{i,j}$ such that $(m-1)\Delta \leq |u_s - t_{i,\ell}| < m\Delta$, thus producing simple inequalities involving the coherence and interference functions.
\begin{lemma}\label{lem:bound_G}Under the assumptions of Theorem~\ref{thm:main-result}, one has
    \begin{subequations}\label{eq:bound_G}
    \begin{align}\label{eq:calc_interference}
    \norm{\bG^i_a}_{\infty} 
    &\leq \I_a(\theta_i^\star, \Delta) + C_\Delta \varepsilon, \\
    \norm{\bG_a^{i^{\top}}\bG_b^j}_{\infty}
    &\leq \C_{a,b}(\theta_i^\star, \theta_j^\star, \Delta) + 2 C_\Delta \varepsilon. \label{eq:calc_coherence}
\end{align}
\end{subequations}
\end{lemma}
Furthermore, from the Lipschitz assumption on the manifold map $\btheta \mapsto \bm{G}(\btheta)$, one can control the residual error as
\(
    \norm{\bm{r}}_\infty \leq \left(\norm{\bm{G}(\btheta^\star)}_\infty  + K \norm{\etab^\star}_\infty \right) \varepsilon.
\)
Thus, by inspection of the block structure, it can be inferred with the triangle inequality that
\begin{equation}\label{eq:bound_offdiag}
    \norm{\nabla^2 \mathcal{L}(\btheta, \etab) - \D}_{\infty} \leq r^\star + q^\star \varepsilon.
\end{equation}
Substituting~\eqref{eq:bound_D} and~\eqref{eq:bound_offdiag} into~\eqref{eq:weyl} yield the main result. \IEEEQEDhere

\section{Numerical Experiments}\label{sec:simulations}

\subsection{Validation of Theorem~\ref{thm:main-result}}

In this section, we numerically validate Theorem~\ref{thm:main-result}. To this end, we reconstruct $p = 2$ modalities within the class of Lorentz PSFs $g(\theta, t) = \theta / (\pi (\theta^2 + t^2))$. The ground truth PSF parameters are $\theta^\star_1 = 2\cdot 10^{-5}$ and $\theta^\star_2 =10^{-3}$. The model orders for each modality are $L_1 = 10$ and $L_2 = 5$ with linear argument $\{\eta^\star_{1,\ell} = \tfrac{1}{L_1}\}_{\ell=1}^{L_1}$, $\{\eta^\star_{2,\ell} = \tfrac{1}{L_2}\}_{\ell=1}^{L_2}$. 

\begin{figure}[t]
    \centering \includegraphics[width=\linewidth]{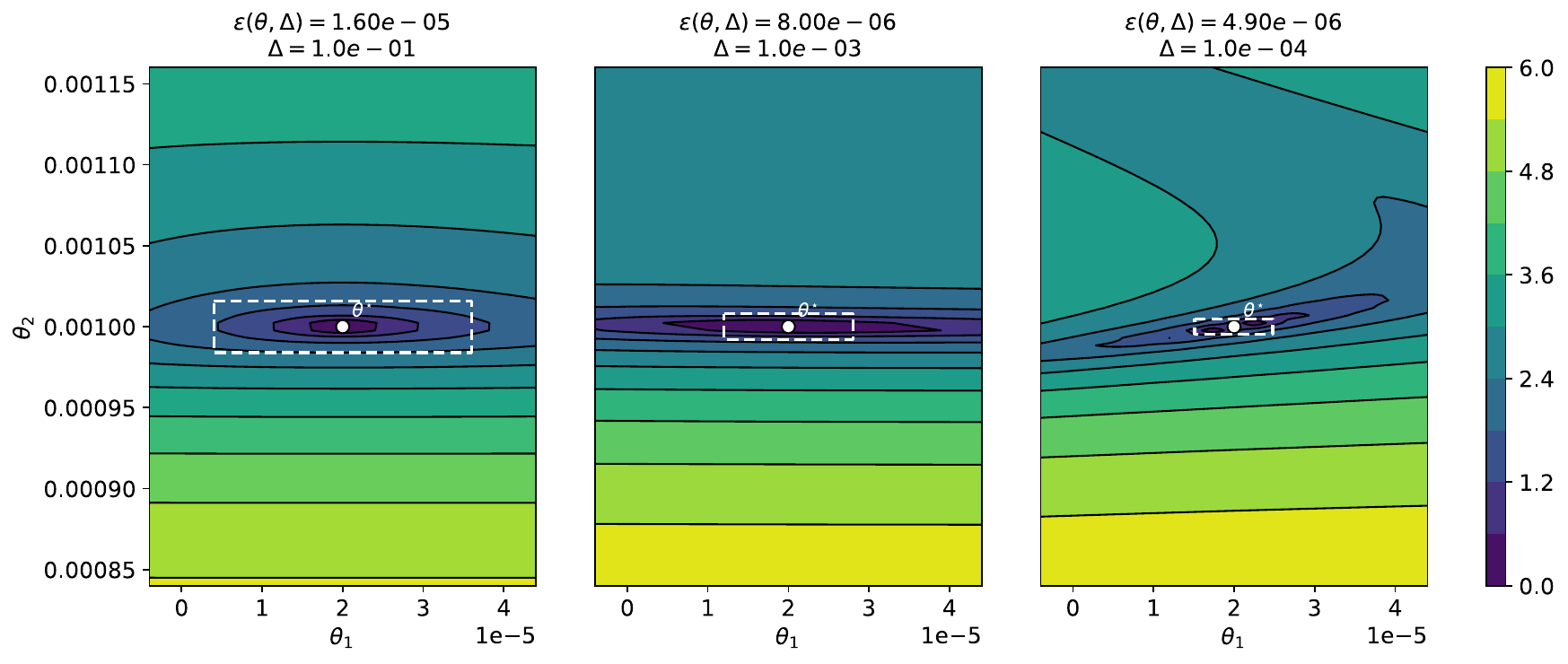}
    \vspace{-12pt}
    \caption{The level sets of the restricted map $\bm{\theta} \mapsto \L(\btheta, \etab^\star)$, and the strong basin radii (in white/dashed) predicted by Theorem~\ref{thm:main-result}, for decreasing separation parameter $\Delta$.}
    \label{fig:basins-theta}
\end{figure}

Since $p=2$, the restricted loss $\bm{\theta} \mapsto \L(\btheta, \etab^\star)$ can be visualized as a contour plot, which is pictured in Figure~\ref{fig:basins-theta} for three different values of $\Delta$. The lower bound on the radius of the basin of attraction guaranteed by Theorem~\ref{thm:main-result} is indicated. It can be appreciated that the radius worsens as $\Delta$ decreases, resulting in a progressively smaller convergence region.

\begin{figure}[t]
    \centering
    \includegraphics[width=\linewidth]{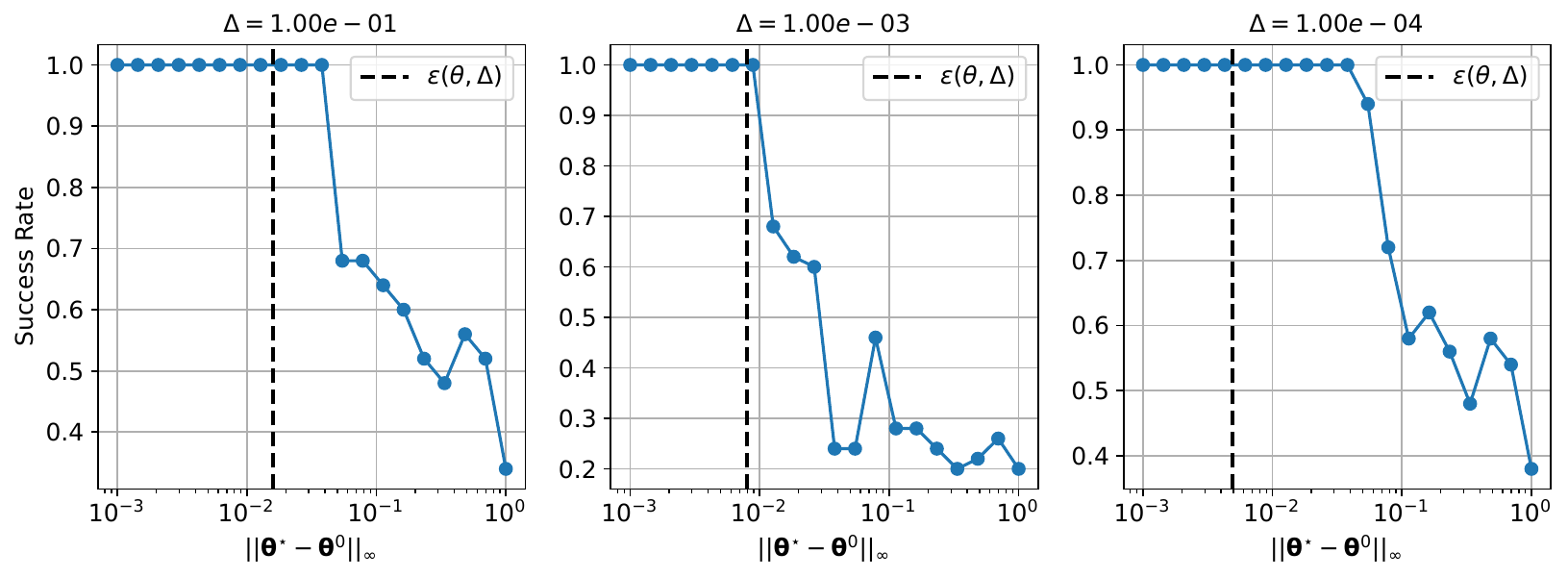}
    \vspace{-12pt}
    \caption{Success rate of \textsc{scipy}'s least squares solver to recover the ground truth $\btheta^\star$ as a function of of the $\ell_{\infty}$-distance between the initial guess $\btheta^{0}$ and the ground truth $\btheta^{\star}$. The results are averaged over 50 trials. The lower bound on the radius of the basin predicted by Theorem~\ref{thm:main-result} is shown in a dashed vertical line.}
    \label{fig:basins-monte-carlo}
\end{figure}

In a second experiment, we challenge the radius $\varepsilon_0$ predicted by Theorem~\ref{thm:main-result} through Monte Carlo simulations. Specifically, we randomize an initialization \(\btheta^0\) and estimate a nonlinear solver's success rate of recovering $\btheta^\star$ as a function of $\|\bm{\theta}^\star - \bm{\theta}^0\|_{\infty}$. The results are reported in Figure~\ref{fig:basins-monte-carlo}. The Monte Carlo simulations converge with empirical probability~$1$ below the threshold predicted by Theorem~\ref{thm:main-result}, which corroborates our result.
We highlight that for $\Delta = 10^{-3}$, our theoretical bound on the basin size is sharp, whereas, for smaller $\Delta$, the bound becomes more pessimistic, owing to the extremely non-convex nature of the problem in that case.

\subsection{Application to Laser-induced Breakdown Spectroscopy}

As a final experiment, we wish to assess the performance of our method on spectroscopical data produced by a LIBS instrument. A spectrum comprises a set of spectral lines with fixed emission wavelengths. Those emission wavelengths are physical constants; hence, they are known and given for each chemical species. However, during acquisition, spectral lines become broadened by various physical interactions in the plasma and vary with the experimental setup, the chemical composition of the sample, and the species. The overarching goal of the analysis of spectroscopical data is recovering the concentration of each element/ion pair present in the sample from the observation of the emission spectrum.

We model the emission spectrum measurement as in Subsection~\ref{subsec:problem_formulation} where $p$ is the number of chemical species to identify in the measured chemical compound, and $\theta_i$ is the shape parameter of the $i$-th species, for $i \in \{1, \dots, p\}$. The species concentration vector $\bnu^\star$ can be modeled as a linear function of the emission amplitudes
\(\etab^\star = \A \bnu^\star
\),
where $\A$ is a matrix which can be estimated a priori.

\begin{figure}[t]
    \centering
    \includegraphics[width=\linewidth]{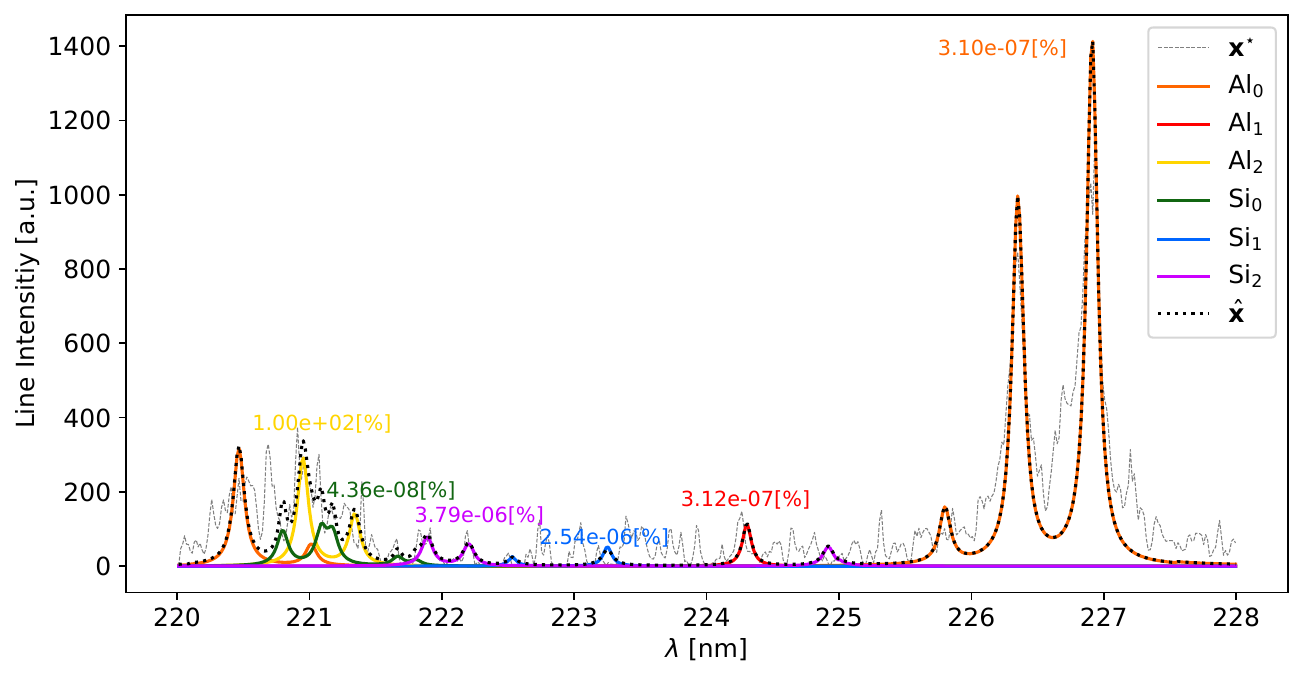}
    \vspace{-14pt}
    \caption{Sampled and fitted LIBS spectrum of an aluminum compound with corresponding ions indicated alongside their estimated concentrations.}
    \label{fig:libs-aluminum}
\end{figure}

We deploy our numerical method on a calibrated metallic sample composed of $97.686\%$ aluminum and $2.313\%$ trace elements, with silicon being the most abundant ($0.896\%$). We aim to estimate the concentration of each aluminum element/ion pair. Since the measurement is performed using a LIBS instrument, the observation model is given by 
\begin{align}
    \bx_{\mathrm{obs}} = \mathbf{\Xi}\left(\bG(\btheta^{\star})\etab^{\star} + \bm{b} + \bm{\xi}\right),
\end{align}
where $\mathbf{\Xi}$ is the apparatus' transfer function, $\bm{b}$ is the spectrum's baseline emission~\cite{baseline-removal} and $\bm{\xi}$ is a noise term, which we have observed follows a $\chi^2$ distribution in practice.
In view of~\eqref{eq:signal-model}, we first perform a preprocessing step to remove the affine term $\mathbf{\Xi}\bm{b}$, resulting in the modified observation $\tilde{\bx} = \bx_{\mathrm{obs}} - \mathbf{\Xi}\hat{\bm{b}}$. We then construct $\bG$ using emission line location information from the NIST database for atomic spectra~\cite{NIST_ASD} and solve
\begin{align}
    \argmin_{\btheta, \etab} \left\{\L(\btheta, \etab) = \frac{1}{2} \norm{\mathbf{\Xi}\bG(\btheta)\etab - \tilde{\bx}}_2^2 \right\}.
\end{align}
Figure~\ref{fig:libs-aluminum} shows the observed and reconstructed emission spectrum, with a fit error of $15.33\%$, while the relative estimation error in the aluminum concentration is $0.023\%$.

\section{Conclusion}\label{sec:conclusion}

In this paper, we formulated a non-convex least square problem to unmix sparse signals with known support that have been convolved with multiple PSFs lying in a univariate manifold. In Theorem~\ref{thm:main-result}, we have provided a lower bound on the radius of the strong basin of attraction of our numerical method in the noiseless case. Immediate extensions include extending the analysis to the presence of noise and considering recovery over a higher dimensional PSF manifold. Furthermore, harnessing sparsity constraints into $\etab$ is a research direction of practical interest to the analysis of spectroscopic data for the purpose of estimating the concentration of chemical species in a compound when its composition is unknown a priori.  

\clearpage
\IEEEtriggeratref{4}
\renewcommand*{\bibfont}{\footnotesize}
\printbibliography

\end{document}